\documentclass[12pt]{article}

\usepackage{amsmath,amssymb}
\usepackage{graphicx}
\usepackage{caption}
\usepackage{color}
\usepackage{dcolumn}
\usepackage{bm}
\usepackage[numbers,super,comma,sort&compress]{natbib}

\newcommand{\trove}{{\sc TROVE}}
\newcommand{\schr}{Schr\"{o}dinger}
\newcommand{\ai}{\textit{ab initio}}

\newcommand{\p}{^\prime}
\newcommand{\pp}{^{\prime\prime}}

\newcommand{\ket}[1]{\vert #1 \rangle  }
\newcommand{\bra}[1]{\langle #1 \vert  }

\newcommand{\Cv}[1]{${\mathcal C}_{#1{\rm v}}$}
\newcommand{\Dh}[1]{${\mathcal D}_{#1{\rm h}}$}
\newcommand{\Ch}[1]{${\mathcal C}_{#1{\rm h}}$}

\newcommand{\Td}{${\mathcal T}_{\rm d}$}

\newcommand{\Cs}{${\mathcal C}_{\rm s}$}

\newcommand{\2}{$_{2}$}
\newcommand{\3}{$_{3}$}
\newcommand{\4}{$_{4}$}

\definecolor{background-color}{gray}{0.98}

\title{The ExoMol project: Software for computing large molecular line lists}
\author {Jonathan Tennyson and  S.N. Yurchenko\\
Department of Physics and Astronomy, University College London,\\ London, WC1E 6BT, UK}
\begin{document}

\maketitle

\begin{abstract}

The use of variational nuclear motion programs to compute line lists
of transition frequencies and intensities is now a standard procedure.
The ExoMol project has used this technique to generate
line lists for studies of hot bodies such as the atmospheres of exoplanets
and cool stars. The resulting line list can be huge: many contain
10 billion or more transitions. This software update considers
changes made to our programs during the course of the project to
allow for such calculations. This update considers three programs:
{\sc Duo} which computed vibronic spectra for diatomics, {\sc DVR3D}
which computes rotation-vibration spectra for triatomics, and
{\sc TROVE} which computes rotation-vibration spectra for general
polyatomic systems. Important updates in functionality include the
calculation of quasibound (resonance) states and Land\'e $g$-factors
by {\sc Duo} and the calculation of resonance states by  {\sc DVR3D}.
Significant algorithmic improvements are reported for both
{\sc DVR3D} and {\sc TROVE}.
All three programs are publically available from
ccpforge.cse.rl.ac.uk.
Future developments are also considered.
\end{abstract}
\clearpage
\section*{Graphical Table of Contents}
Molecular spectra provide important remote sensing fingerprints. However
hot molecules can undergoing very large numbers of possible transitions: billions for even fairly small molecules such as methane.  
Nuclear
motion software  based on the use of the
variational principle used to compute line lists is discussed
and the adaptation of the programs  to the demands of 
computing huge lists of molecular transitions
described.
\setcounter{figure}{-1}
\begin{figure*}[h!]
\includegraphics[height=100mm]{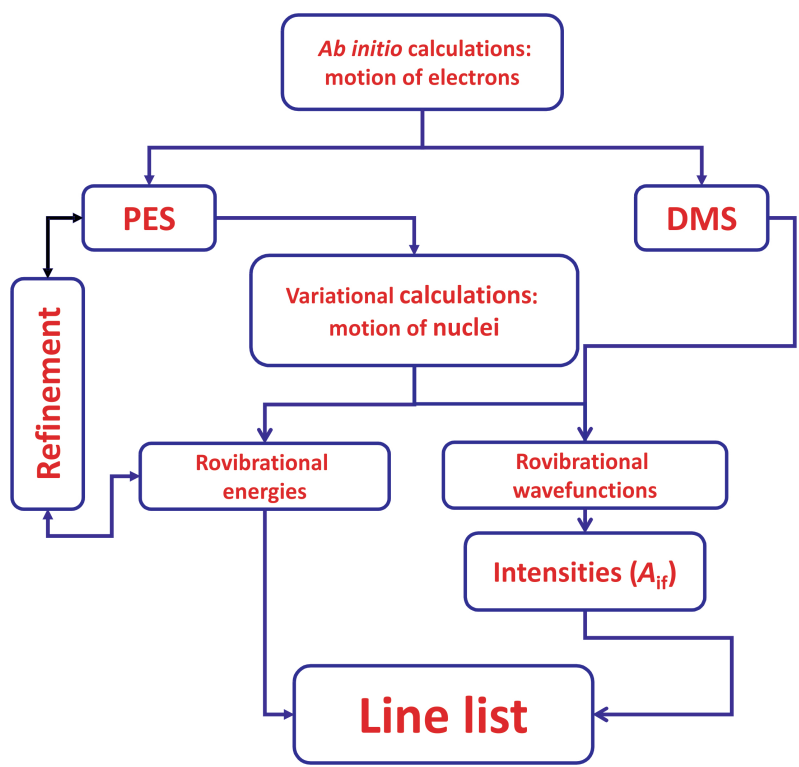}
\caption{Work flow illustration the
technique used to compute a  line list of molecule transitions.  }
\end{figure*}

\clearpage
\section {Introduction}

The ExoMol projects aims to compute line lists of molecular transitions
which are important for the study of hot atmospheres, particularly those
of exoplanets, brown dwarfs and cool stars.\cite{jt528} In practice
these line lists are also useful for a variety of terrestrial applications
as well as for models of non-thermal environments such as masers.
The project has produced comprehensive line lists for a number of molecules
including
BeH,
MgH and
CaH \cite{jt529},
SiO \cite{jt563},
HCN/HNC \cite{jt570},
CH$_4$ \cite{jt564},
NaCl and
KCl \cite{jt583},
PN \cite{jt590},
PH$_3$ \cite{jt592},
H$_2$CO \cite{jt597},
AlO \cite{jt598},
NaH \cite{jt605},
HNO$_3$ \cite{jt614},
CS \cite{jt615},
CaO \cite{jt618},
SO$_2$ \cite{jt635},
HOOH \cite{jt620},
H$_2$S \cite{jt640},
SO$_3$ \cite{jt641},
VO \cite{jt644},
H$_3^+$ \cite{jtH3+}
and CrH \cite{jtCrH};
the diatomic studies generally include consideration
of all important isotopologues. These line lists are large with,
for example, the line list for the diatomic $^{40}$Ca$^{16}$O containing
over 28 million transitions \cite{jt618}, and those
for the polyatomic systems CH$_4$, PH$_3$, H$_2$CO, HOOH and SO$_3$
containing 10 billion or more lines. These calculations can also be
used for other purposes such computing radiative
lifetimes of individual states\cite{jt624} and
thermally-averaged properties.

Computing these line lists has led us to develop or improve specialist
programs designed to study the nuclear motion problem of the various
molecules under consideration.  This software update describes these
developments.  A common theme of all these programs is the direct
solution of the nuclear motion Schr\"odinger equation using a
variational treatment.  In the next section we outline the overall
methodological approach adopted by ExoMol. In the following sections
we consider the main programs used under the project. They are grouped
by the type of system studied. Section 3 considers diatomic systems,
for which we use Le Roy's program {\sc Level} \cite{lr07} and our
program especially developed for the project, {\sc Duo}.\cite{jt609}
Unlike the other programs considered here, {\sc Duo} is designed for
the calculation of vibronic spectra and can treat problems involving
coupled potential energy curves. Section 4 considers triatomic systems
for which the exact kinetic energy nuclear motion code {\sc DVR3D}
\cite{DVR3D} has been employed. For tetratomic systems calculations
have largely been performed with {\sc TROVE} \cite{07YuThJe.method}
although {\sc WAVR4} \cite{jt339} has also been tested.\cite{jt553}
{\sc TROVE}, which has also been used to study methane, will be
considered in section 5.  Finally a new hybrid methodology based on
the combined use of the variational principle and perturbation theory
has been developed for larger systems. This will be considered in
section 6. The final section considers our conclusions and prospects
for the future.

A number of other groups are involved in projects computing extensive
molecular line lists for astronomical or other purposes, again largely
using especially developed software for the nuclear motion problem.
These include the NASA Ames group of Huang, Schwenke and Lee who use
the polyatomic nuclear motion program {\sc VTET}
\cite{96Schwen.method} and have also been undertaking theoretical
developments.\cite{15Schwenke.diatom} Tyuterev and Rey from the
University of Rheims in collaboration with Nikitin from the Tomsk
Institute of Atmospheric Optics have computed line lists for a number
of polyatomic species\cite{14ReNiTy.CH4,TheoReTS} using either their variational polyatomic code
or contact transformation approach. Bowman's group has developed a
very efficient general approach to compute ro-vibrational intensity of
polyatomic molecules using MULTIMODE \cite{MULTIMODE} which has also
been used to generate hot line lists.\cite{09WaScSh.CH4} Finally, we note
that Bernath's group has produced a number of diatomic line lists
based on the use of {\sc level} with the effects of electron spin
treated using perturbation theory
\cite{14BrBeWe.NH,14MaPlVa.CH,14RaBrWe.CP}.  We note that there are
also a number of studies which are the product of collaboration
between the various groups.\cite{jt583,jt635,14BrRsWe.CN}

\section{Method}

\begin{figure}
\includegraphics[height=100mm]{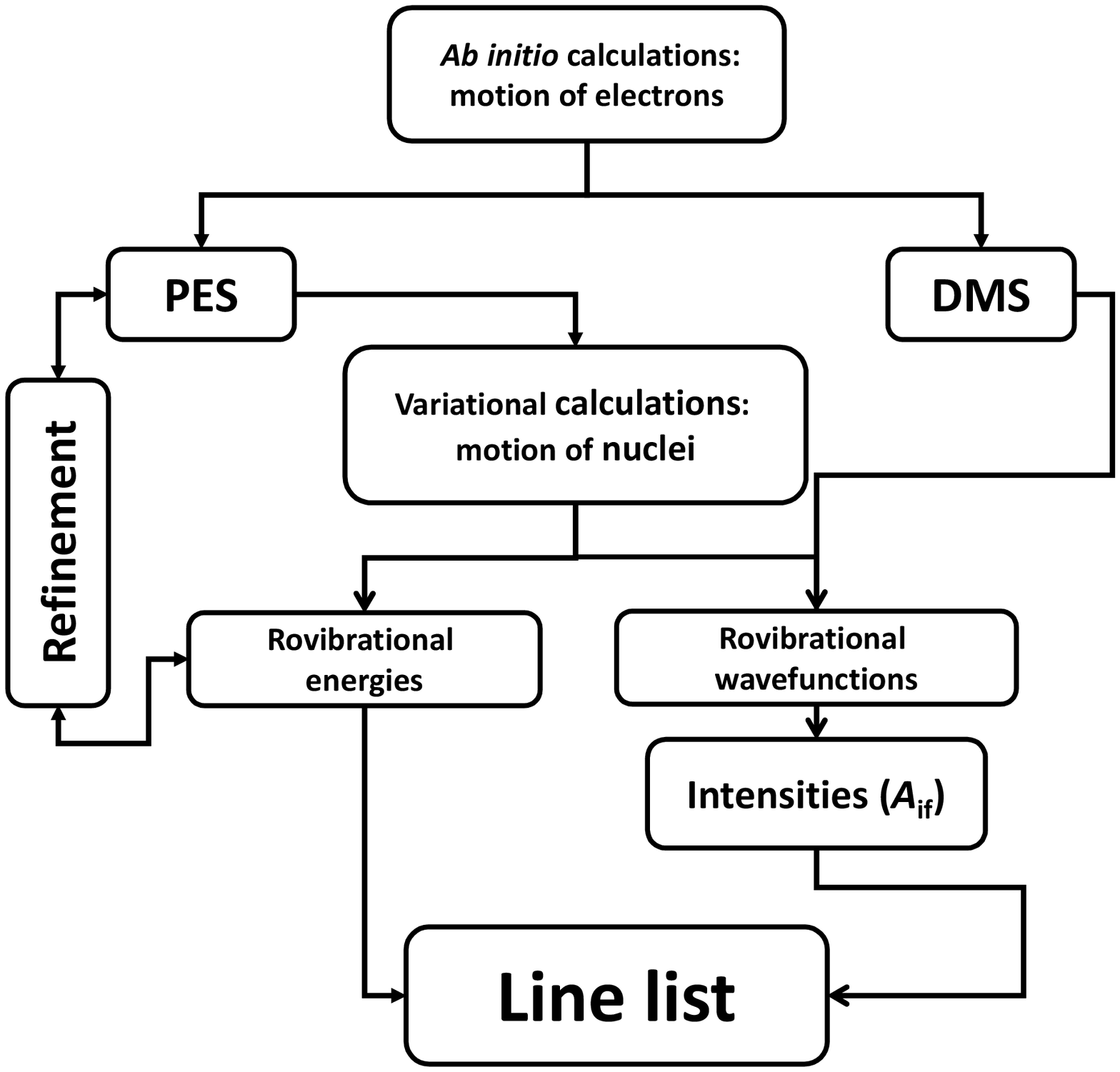}
\caption{\label{f:workflow} A typical work flow of the line list production.  }
\end{figure}

The general methodology used by us for constructing line lists has been
extensively discussed elsewhere; in particular Lodi and Tennyson \cite{jt475}
gave an introduction on how to perform such calculations and Tennyson
\cite{jt511} reviewed the methodology used by the ExoMol project which
is summarised in
Fig.~\ref{f:workflow}.

The first step in the calculation is construction of a potential
energy surface (PES) using a high-level, {\it ab initio} procedure,
which we generally do with {\sc MOLPRO} \cite{12WeKnKn.methods}. At
the same time the computation of the appropriate dipole moment
surfaces (DMS) is performed. These then form the input to the
appropriate nuclear motion program; these programs are the focus of
the present article. In only a very few cases
\cite{jt298,jt347,jt506,jtH3+} are completely {\it ab initio}
procedures the best choice for obtaining an accurate line list.  In
general this is only true for systems with very few electrons.
Otherwise it is necessary to refine the calculation using experimental
data.

There are three methods of improving the calculated line list on
the basis of empirical data. The most common one is to refine the PES
in the manner illustrated in Fig.~\ref{f:workflow}. This
methodology, which is widely used by a number of groups,
\cite{MORBID,ps97,01TyTaSc.H2S}
involves either adjusting
parameters in the original fit of the PES or adding an auxiliary function
which captures changes to this PES. The second method, which can only
be used in programs such as {\sc TROVE} which uses uncoupled vibrational
and  rotational basis functions, the so-called $J$$=$$0$ representation
of rotational excitation \cite{jt466}, involves band origin shifts.
In this method, the vibrational band origins that are computed in
the rotationless ($J$$=$$0$) step of the calculation  can be shifted
to the observed one prior to solving the fully-coupled rotation-vibration
problem \cite{jt466}. The third method involves substituting
empirical energy levels at the end of the calculation. The format used
for storing ExoMol line lists \cite{jt548,jt631} involves creating
a states file which contains all energy levels and associated quantum
numbers. There are now well-established procedures for extracting
experimentally-determined energy levels from high resolution spectra
\cite{jt412,12FuCsi.method,jt562,jt637} and these energies can simply
be used to replace the computed ones in the states file.

The situation with DMS is very different. The evidence is that
DMS can be calculated {\it ab initio} more accurately than they
can be obtained by inverting experimental data \cite{jt156}. Furthermore
theoretical procedures have been developed which allow the assignment
of uncertainties to individual
transition intensities \cite{jt522,jt625}, although at present these
are too onerous to be used routinely for the very large line lists being
considered here. Reviews discussing the theoretical determination
of accurate DMS have been given by each of us \cite{13Yuxxxx.method,jt573}.

The nuclear motion programs which are the focus of this software review can
be thought of as solving the Schr\"odinger equation implied
by the nuclear-motion Hamiltonian:
\begin{equation}
\hat{H} = \sum_I \frac{-\hbar^2}{2 M_I}\nabla_I^2 + V(\underline{Q}),
\end{equation}
where $I$ runs over the $3N$
coordinates of the $N$ nuclei, each of mass $M_I$, and $V(\underline{Q})$
is the  PES expressed in internal coordinates $Q$.
Of course this expression already assumes the Born-Oppenheimer approximation
and neglects any couplings between PESs. To make progress with
Hamiltonian (1) it is necessary to separate out the centre-of-mass coordinates
which represent the translation motion of the whole molecule. The
methods below also all work in body-fixed coordinates which involve
separating the rotations from the vibrations by fixing an axis system
to the molecular frame and using internal coordinates to express the
vibrational coordinates. Precisely how this is done varies between
the different programs.

In the standard variational approach the energies and wave functions
of the body-fixed Hamiltonian are obtained by using appropriate basis
functions to represent the rotational and vibrational motion, and then
diagonalizing the resulting matrix; see, for example, the
review by  Bowman {\it et al.} \cite{08BoCaMe.methods}.
For the rotational motions the
choice of basis functions is straightforward as symmetric top eigenfunctions
(Wigner $D$-matrices) form a complete set. For the vibrational
coordinates it is often preferable to use a grid-based discrete
variable representation (DVR) \cite{lc00} rather than actual
functions. Furthermore, rather than diagonalizing the resulting
Hamiltonian matrix in a single step, our approach often uses
intermediate diagonalizing steps so that the final matrix diagonalization
is as compact as possible.

Ones ability to diagonalize the large matrices necessary for obtaining
the many energies and wave functions required for computing hot line lists
usually provides the computational bottleneck in these calculations.
However, the very large number of transition probabilities, which we
generally choose to represent as Einstein A coefficients, that need to be
computed usually means that this step can come to dominate the
actual computer
time used. Measures to mitigate this are discussed below.

\section{Diatomics systems}

Le Roy's program {\sc LEVEL} \cite{lr07} is our choice for computing the spectra
of closed shell ($^1\Sigma$) diatomic molecules. The program has
been refined over many years by Le Roy and has been used by us without
further changes.

However, for more complicated diatomic systems, in particular ones
involving coupled electronic states or non-$\Sigma$ states, we have
developed our own program, {\sc Duo}. A first release of {\sc Duo} has
just been published \cite{jt609} and the reader is refered to this
paper and an associated topical review \cite{jt632} for full details.
{\sc Duo} is still under regular development and a number of
improvements to the functionality of the published version
have been made of which we
highlight three here.

First, the published version of {\sc Duo} only considers truly bound
states. However, there are a number of situations where it is necessary
to consider quasi-bound or resonance states, or indeed the continuum
itself. Shape resonances arise when rotational excitation leads
to quasi-bound states being trapped by the centrifugal barrier. There
are also Feshbach resonances which undergo
pre-dissociation caused by coupling to dissociative states. Finally,
it is sometimes necessary to consider
spin-orbit effects on bound states caused by coupling to either resonances
or the continuum.
A facility has been added to {\sc Duo} which allows an artificial wall to be
placed at large internuclear separation; this has the effect of
discretizing the continuum and allowing localized, resonance states
to be identified \cite{jt643}.

Second, the energy levels of open shell molecules are sensitive to the
effects of magnetic fields. The behaviour of molecules in a magnetic
field provides a spectroscopic tool as well as being important in fields
as diverse as molecular trapping \cite{14BaMcNo.diatom} and astrophysics \cite{02BeSoxx.diatom}.
For an open shell diatomic, the splitting, $\Delta E_{JM}$, due to weak magnetic field of strength
$B$, otherwise known
as the Zeeman splitting, is given by
\begin{equation}
 \Delta E_{JM} = g_J M \mu_B B ,
\end{equation}
where $J$ is the total angular momentum quantum number and $M$ is its projection along the direction
of the magnetic field. Here $g_J$ is the effective Land\'e g-factor for the given level and $\mu_B$ is the Bohr magneton. In simple cases
$g_J$ can be evaluated
in the Hund's case (a) basis used by {\sc Duo}, using the expression
\begin{equation}
 g_J = \sum_n |C^{J,\tau}_{\lambda,n}|^2  \frac{\Lambda_n g_L+ 2\Sigma_n g_S}{J(J+1)}, \label{eq:gJ}
\end{equation}
where $g_L=1$ and $g_S=2.0023$ are the orbital and spin $g$-factors
respectively. In eq.~(\ref{eq:gJ}),
 $C^{J,\tau}_{\lambda,n}$ is the eigenvector of the $\lambda^{\rm th}$ state with
good quantum numbers $J$ and $\tau$ (parity) and $n$ is a compound basis index
\begin{equation}
 \ket{n} = \ket{{\rm state},J,\Omega,\Lambda,S,\Sigma,v}.
\end{equation}
In this `state' denotes an electronic state with electron spin $S$ and
$v$ an associated vibrational state. $\Omega (=\Lambda +\Sigma)$,
$\Lambda$ and $\Sigma$ are projection of the orbital angular momentum
and spin, $\boldmath{S}$, onto the body-fixed molecular axis,
respectively. In Eq.~(\ref{eq:gJ}), $\Lambda$ and $\Sigma$ are
subscripted by $n$ to emphasize that they are not conserved quantities
but their value depends on the state part of the basis.  Berdyugina
and Solanki\cite{02BeSoxx.diatom} give a more complete expression for
$g_J$ which allows it to be evaluated correctly using {\sc Duo}
wavefunctions even when the molecule is not well-represented by Hund's
case (a).  An extension to {\sc Duo} to evaluates this general
expression has recently been written and tested for a few diatomic
systems, notably CrH, C$_2$ and AlO.\cite{jtgfac}

Finally, visualization of wave functions can be very helpful for
interpreting results. The latest version of {\sc Duo} has incorporated
plotting routines to aid the inspection of the results.

\section{Triatomic systems}

The {\sc DVR3D} program suite obtains variationally exact solutions
for the bound-state, three-atom nuclear motion problem for a given PES
within the Born-Oppenheimer approximation. The program has been developed
over a number of years originally starting as a finite basis set
procedure \cite{jt20,jt48,jt79,jt128} before evolving \cite{jt129} to
one which is based on the use of DVR in all vibrational coordinates
\cite{jt130,jt160}.  DVR3D and its predecessors have been benchmarked
against other, similar, nuclear motion codes such {\sc VTET},
and indeed {\sc TROVE},
to confirm the accuracy
of both the computed vibration-rotation energy levels
\cite{jt309,jt635} and transition moments \cite{jt78}.

The current published release of the {\sc DVR3D} program suite
\cite{DVR3D} is actually the third but dates back to 2003. Since then
{\sc DVR3D} has undergone a large series of developments, not least to
facilitate the calculation of huge line lists.  All modules have also
been subject to a re-write to both make them more consistent and to
bring the programming up to a more modern programming standard.

\begin{figure}
\centering
\includegraphics[height=100mm]{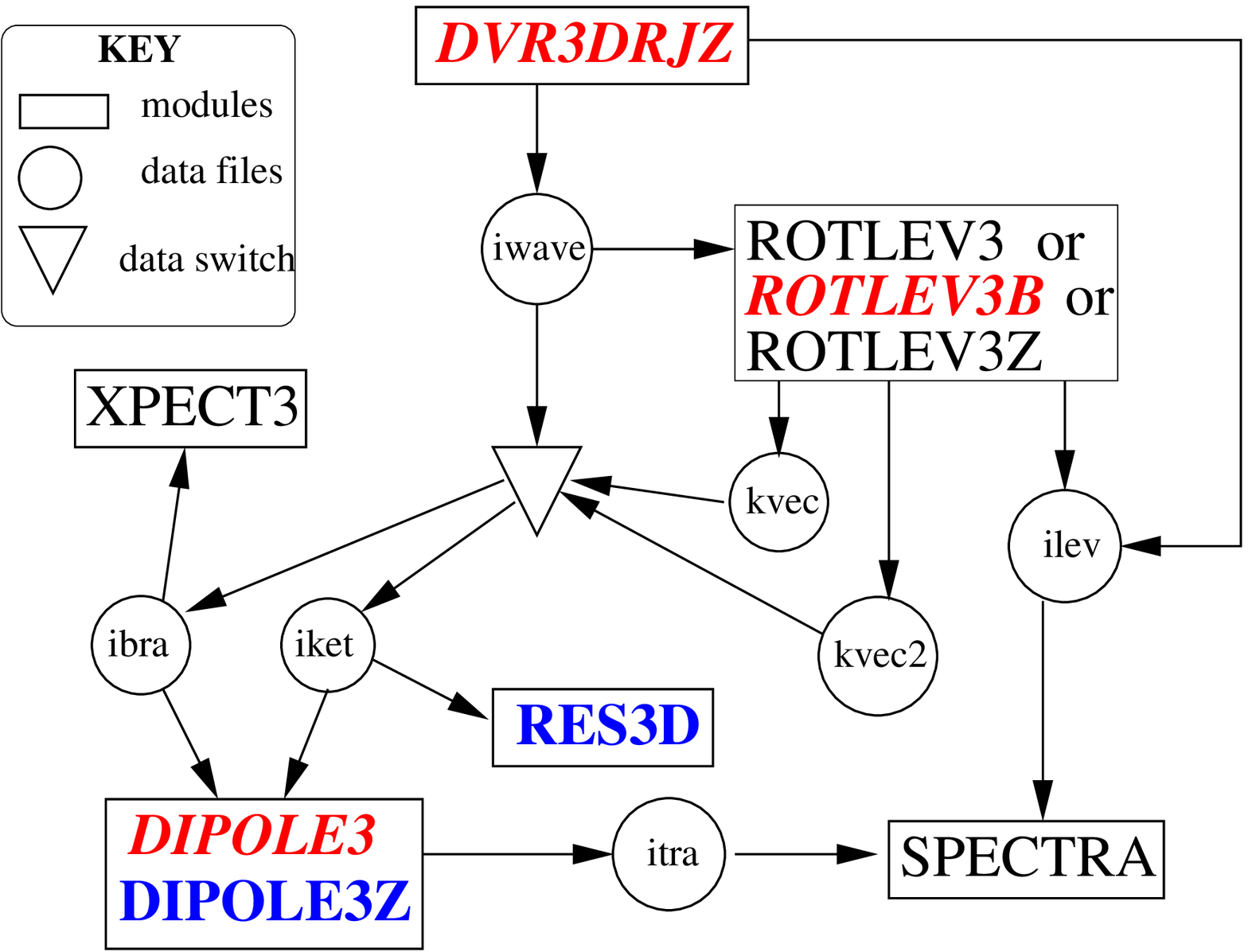}
\caption{\label{f:dvr3d} Flow chart illustrating the various modules of program
{\sc DVR3D}. Molecules in bold are new and those in italic have significant
algorithmic improvements since the last published release of the code \cite{DVR3D}.}
\end{figure}

Figure~\ref{f:dvr3d} gives the flow structure for the current version
of {\sc DVR3D}. The main driving module, {\sc DVR3DRJZ}, solves the
vibration-only or Coriolis-decoupled vibration-rotation problem.  For
rotationally-excited molecules, the results of {\sc DVR3DRJZ} provide
the basis functions used in one of the {\sc ROTLEV} modules to solve
the fully-coupled vibration-rotation problem; where the choice of
module depends on the axis embedding used. The solutions to
vibration-rotation problem can be used to compute expectation values
of given variable in {\sc XPECT}, this module is particularly used in
performing fits of the PES to spectroscopic data where the
Hellmann-Feynman theorem can be used to evaluate the expectation
values of derivatives of the PES with respect to parameters of the
fit. These same vibration-rotation eigenvectors can be used to compute
line strengths in {\sc DIPOLE} which in turn provides the necessary
information to generate spectra. The new modules and significantly
amended modules have been highlighted in this figure and are
discussed below.

A new module, {\sc RES3D}, has been written \cite{jt443} which can be
used to characterize quasibound or resonance states lying above
dissociation.  The automated procedure for doing the analysis has been
successfully used to study resonances in both H$_3^+$ \cite{jt443} and
water \cite{jt500}. Details of how this module works are given below.

In addition the functionality of {\sc DVR3D} has been increased by a
thorough re-write of the codes in which the $z$-axis is placed
perpendicular to the plane of the molecule ($z$-perpendicular
embedding option \cite{jt290,jt310}).  This option is useful for
molecules, such as H$_3^+$, whose projected rotational motion is
usually quantised along this axis.  A new module, {\sc DIPOLE3Z}, is
introduced which computes transition dipoles for the $z$-perpendicular
embedding case.

Algorithmic improvements include the following:
\begin{itemize}
\item The automated Gauss-(associated) Legendre quadrature generation
  procedure, which was adapted from one given by Stroud and Secrest
  \cite{jt66} has been replaced by a brute force one which involves
  finding zeros in the polynomial equation \(P_N^k(x)=0\) for the $N$
  point quadrature.  This was found to be essential for grids with $N
  > 90$ and has been successfully used for $N$ up to 150 \cite{jt635}.
  The automatic check on the validity of grid obtained by comparing
  summed weights with the analytic value given by Stroud and Secrest
  has been retained.
\item For large calculations, module {\sc ROTLEV3b}
in  the published version of {\sc DVR3D} can
spend a long time constructing the final Hamiltonian matrix.
{\sc ROTLEV3b} uses vibrational functions
generated in the first step of the calculation \citep{jt46} to provide
basis functions for the full ro-vibrational calculation performed by
{\sc ROTLEV3b}. For high $J$ calculations this algorithm involves
transforming large numbers of off-diagonal matrix elements to the
vibrational basis set representation, see Eq.~(31) in \citet{jt114}.
This step has been re-programmed as two successive summations rather
than a double summation at the cost of requiring an extra, intermediate
matrix \cite{jt640}. This had the effect of reducing the cost of Hamiltonian
construction to below that of Hamiltonian diagonalization, which is
generally the case for the other modules of {\sc DVR3D}.
\item {\sc Dipole3}  by default computes all
  transition dipoles between the bra and ket wave functions it is asked
  to process. For large line lists, computing transition dipoles
  actually dominates computer usage and this can be inefficient.
  These line lists are usually characterized by a lower energy cut-off,
  which determines the temperature range for which the line list is
  valid, and an upper energy cut-off which determines the frequency
  range. Computing transition moments between these ranges is
  expensive and unnecessary. New input variables have been introduced
  to avoid this \cite{jt538}.
\item {\sc DVR3DRJZ} employs an algorithm which relies on solving a Coriolis-decoupled
  vibrational problem for each $(J,k)$, where $k$ is the projection of
  the rotational angular momentum onto the chosen body-fixed $z$ axis and
  $J$ is the rotational motion quantum number. This provides a basis set from which functions
  used to solve the fully-coupled ro-vibrational problem are selected
  on energy grounds \cite{jt66}.  Hot line list can involve
  calculations with high $J$ and experience has shown that in this
  case not all $(J,k)$ combinations are actually needed.  An option
  has been implemented where unneeded high $k$ calculations are not
  performed \cite{jt635}. In practice, this does not save much
  computer time, since the initial $(J,k)$ calculations are quick, but
  does save disk space.
\item Again for large calculations, the
algorithm used by module {\sc DIPOLE3} to read in the wave functions
required a lot of redundant reads. {\sc DIPOLE3} has been re-structured to
reduce the number of times the wave functions need to be read \cite{jt635}.
\end{itemize}

Finally, matrix
diagonalization is the rate-limiting step in most applications
of {\sc DVR3D}. A number of new real, symmetric matrix diagonalizers
have been added to the  LAPACK software package \cite{99AnBaBi.method}.
The diagonalizers implemented in {\sc DVR3D} have been
changed where appropriate.

\subsection{\label{sec:cap_method} Resonance detection}

Resonances can be detected by the behaviour of states lying in the
continuum upon the introduction of a complex absorbing potential
(CAP). To do this
the dissociating system's PES is augmented with a
complex functional form that absorbs the continuum part of the
wave function. This non-Hermitian Hamiltonian produces
$L^2$ wave functions above the dissociation threshold that represent
the resonant states in question \cite{93RiMexx.method}.

Formally, an imaginary negative potential that acts on the
dissociation coordinate, $R$, is added to the system's Hamiltonian,
$\hat{H}$:
\begin{equation} \label{eq:CAP}
\hat{H}'= \hat{H} - i\lambda W(R)\ \ ,
\end{equation}
where $\lambda$ is a parameter used to control the CAP's
intensity. The resulting non-Hermitian Hamiltonian, $\hat{H}'$,
defines the energy of the $n^{\rm th}$ resonance, $E_n$, its width,
$\Gamma_n$, and the corresponding $L^2$ wave function, $\Psi_n$,
through the relationship:
\begin{equation} \label{eq:resonance}
  \hat{H}'(\lambda)\Psi_n(\lambda)= \left(E_n(\lambda) - i\frac{\Gamma_n(\lambda)}{2}\right)\Psi_n(\lambda)\ \ .
\end{equation}
To solve Eq.~(\ref{eq:resonance}), $\hat{H}'$ can be projected on a suitable
basis set and diagonalized. In the infinite basis set limit, the
eigenvalues corresponding to the resonant states will be found in the
limit where $\lambda \rightarrow 0$. Fortunately the use of a finite
basis set is both necessary and beneficial: the error introduced by
the CAP and the finite basis set have opposite phase. This implies
that these errors will cancel each other out at some optimal value,
$\lambda_{op}$, thus yielding the complex ``observables'' associated
with the resonant state.

The wavefunctions, $\Psi_n(\lambda)$, satisfying
eq.~(\ref{eq:resonance}) are naturally complex. This represents both
transmission and reflection at the CAP. Theoretically it is best to
minimize reflection \cite{95RiMexx.method}, which can be done
by judicious choice of absorbing potential \cite{02Manolo.method}.
However, our calculations did not find much sensitivity to actual
choice of CAP used.\cite{jt443,jt494} However, the width in particular
is found to be sensitive to convergence of the basis set representation
employed.\cite{jt305}

A search for $\lambda_{op}$ is made by studying the behaviour of the
complex eigenvalues of Eq.~(\ref{eq:resonance}) with values of
$\lambda$ ranging from zero to a large arbitrary value. This results in
$N$ trajectories in the complex plane, each associated with an
eigenvalue $E_n - i\Gamma_n/2$.  Through graphical analysis of these
trajectories it is possible to identify the point in the complex plane
that corresponds to the optimal value $\lambda_{op}$, and hence
estimate the value for the position, $E_n$, and with, $\Gamma_n$, of
the resonant state. This graphical method consists of locating cusps,
loops and stability points in the eigenvalue trajectory, which are
known to occur in positions around the true eigenvalue for the
resonances on the complex plane.\cite{jt443,81MoFrCe}

The approach taken in the new {\sc RES3D} module of  {\sc DVR3D} is to first diagonalize $\hat{H}$ of
the system under study and store the basis elements $\phi_i$, and
eigenvalues $\varepsilon_i$ lying near dissociation.  As one
can expand the functions $\Psi_n$ of Eq.~(\ref{eq:resonance}) onto the basis
set obtained from the bound state calculation:
\begin{equation} \label{eq:wave}
\ket{\Psi_n(\lambda)} = \sum_i c^i_{n}(\lambda)\ket{\phi_i}\ .
\end{equation}
The coefficients $c^i_n(\lambda)$, the resonance energies
$E_n(\lambda)$ and the resonance widths $\Gamma_n(\lambda)/2$ can then
be obtained by diagonalizing the Hamiltonian:
\begin{equation} \label{eq:capdiag}
H'_{ji}= \bra{\phi_j}\hat{H}'\ket{\phi_i}= \varepsilon_i\delta_{ji} - \imath \lambda
\bra{\phi_j}W(R)\ket{\phi_i} \ ,
\end{equation}
where $\varepsilon_i$ is the $i^{th}$ eigenvalue and $\phi_i$ is the
$i^{th}$ eigenvector obtained from the diagonalization of $\hat{H}$.
For systems with many bound states, wave functions associated with the
strongly bound states are not needed when diagonalizing the Hamiltonian so can
be dropped. This means that the  Hamiltonian matrix $H'$ is small, easy
to construct and
cheap to diagonalize which is important as the graphical method relies
on many diagonalizations with different values if $\lambda$.

The $H'$ matrix is complex symmetric matrix which therefore yields the
complex eigenvalues needed to characterize both the position and width
of the resonance. {\sc RES3D} uses LAPACK \cite{99AnBaBi.method}
routine {\sc zgeev} to perform this diagonalization.

\section{Polyatomic systems}

Code {\sc WAVR4} \cite{jt339} provides tetratomic implementation of
the DVR-style approach employed in DVR3D. \cite{jt346} {\sc WAVR4} has
been tested\cite{jt553} against the alternative polyatomic code {\sc
  TROVE}, described below, and found to considerably slower.  There
are a number of reasons for this. Firstly, DVR methods are diagonal in
the potential and coupling appears through the kinetic energy
operator.  Although it is possible to formulate a DVR in general
coordinates\cite{jt91}, this is not efficient as it is only in
orthogonal or polyspherical coordinates \cite{09GaIuxx.method} in which
the kinetic energy operator has a simple form that can be efficiently
evaluated. Secondly, the $J=0$ representation as implemented in {\sc
  TROVE} and discussed below has proved highly efficient for
calculations on rotationally excited polyatomics. This form is not
supported by  {\sc WAVR4} which employs theory which naturally samples
linear geometries \cite{jt312,jt346} where the  $J=0$ representation fails.

\trove\ is a variational method with an associated Fortran~2003 program to
construct and solve the ro-vibrational \schr\ equation for a general polyatomic
molecule of arbitrary structure \cite{TROVE}. The kinetic energy
operator is constructed as an expansion in terms of internal
vibrational coordinates with the expansion coefficients obtained
numerically on-the-fly.  The energies and eigenfunctions  obtained
via a variational approach
can be used to model absorption/emission intensities (absorption
coefficients) for a given temperature as well as to compute temperature
independent line strengths and Einstein coefficients
\cite{05YuThCa.method}. The latter is then used as the input to
construct molecular ExoMol line lists. \trove\ provides an integrated
facility for refining the \ai\ PES in the appropriate analytical
representation \cite{11YuBaTe.NH3}. Being a general program \trove\
requires modules for each molecular type with all individual
specifications including descriptions of the molecular structure,
internal coordinates, and their symmetry properties. \trove\ uses a
symmetry adapted product-type basis set representation with an
automatic symmetrization procedure \cite{16YuYaxx.method}. The
Hamiltonian matrix constructed by \trove\ is factorized into symmetry
blocks corresponding to different irreducible representations. The
molecular symmetry group \cite{04BuJexx.method} is used to classify
the symmetries of the basis and wave functions.
The construction of the ro-vibrational basis set is performed
in three steps: (i) the 1D basis set functions are
obtained either as numerical solution of 1D \schr\ equations using the
Numerov-Cooley method \cite{23Nuxxxx.method,61Coxxxx.method} or the
harmonic oscillator wavefunctions; (ii) \schr\ equations are solved
for reduced Hamiltonians for different types of degrees of freedom
connected by symmetry transformations in order to obtain a more
compact, contracted basis set; the eigenfunctions of the $J$$=$$0$ \schr\
equation are then contracted and used to form the final ro-vibrational
basis set in the $J$$=$$0$ representation \cite{jt466}; (iii) the
final step involves constructing and diagonalizing the symmetrized
ro-vibrational Hamiltonian matrix.

Apart from computing energies and spectra for a series of polyatomic molecules, the program \trove\ has being applied to study some of their properties, for example the so-called rotational energy clustering\cite{09YuOvTh.SbH3,jt580} or the temperature-averaged nuclear spin-spin matrix elements\cite{10YaYuPa.NH3} and isotropic hyperfine coupling constant\cite{15AdYaYuJe.CH3}.

Even prior to the ExoMol project a number of modifications had been
implemented to \trove\ subsequent to its original publication \cite{TROVE}, which have proved to be important for the project. These include:
\begin{enumerate}
  \item Symmetry adapted basis set and contraction scheme based on the reduced Hamiltonian problems  (to be reported soon \cite{16YuYaxx.method});
  \item Intensity calculations \cite{jt466};
  \item $J$$=$$0$ representation \cite{jt466};
  \item Thermal averaging using an expansion of the matrix exponent \cite{10YaYuPa.NH3};
  \item Empirical band center corrections \cite{jt466}.
\end{enumerate}

 The typical \trove\ intensity project consists of the following steps:

\begin{enumerate}
  \item Expansion of the Hamiltonian operator (generating kinetic and potential energy expansion coefficients numerically on-the-fly) as well as of any `external' function (e.g. dipole moment, polarizability, spin-spin coupling or any other property; PES correction $\Delta V$ used in the refinement process \cite{11YuBaTe.NH3});
  \item Numerov-Cooley solution of the 1D \schr\ equations;
  \item Eigen-solutions of the reduced Hamiltonian problems; \label{i:reduced}
  \item Symmetrization of the contracted eigenfunctions from Step~\ref{i:reduced} and construction of the symmetry-adapted vibrational basis set $\phi_{i}$;
  \item Calculation of the vibrational matrix of the Hamiltonian operator as well as external functions (e.g. dipole) when required; \label{i:vib}
  \item Diagonalizaitons of the $J$$=$$0$ Hamiltonian matrices for each irreducible representation in question;
  \item Conversion of the primitive basis set representation (vibrational matrix elements from Step~\ref{i:vib}) to the $J$$=$$0$ representation;\label{i:conv}
  \item Construction of the symmetry-adapted ro-vibrational basis set as a direct product of the $J$$=$$0$ eigenfunctions and rigid rotor wavefunctions;
  \item Construction of the ro-vibrational Hamiltonian matrices for each $J\ge 0$ and irreducible representation $\Gamma$;
  \item Diagonalization of the Hamiltonian matrices and storing eigenvectors for the postprocessing (e.g. intensity calculations) if necessary; \label{i:diag-rovib}
  \item For the intensity calculations (line list production), all pairs of the ro-vibrational eigenvectors (bra and ket) from Step~\ref{i:diag-rovib} (subject to the selection rules as well as to the energy, frequency and $J$ thresholds)  are cross-correlated with the dipole moment $XYZ$ components in the laboratory-fixed frame via a vector-matrix-vector product, where the body-fixed $xyz$ components of the dipole moment from Step~\ref{i:vib} are transformed to the $XYZ$-frame using the Wigner-matrices. \label{i:int}
\end{enumerate}

The main challenge of the ExoMol project is that very high rotational and
vibrational excitations are needed for for accurate descriptions of high-temperature molecular spectra.
This in turn requires larger basis sets and therefore larger Hamiltonian matrices, which associated increase of
the calculation costs in terms of memory (both RAM and storage) and time. For
example, for the SO$_3$ line list \cite{jt641} with extremely high
rotational excitations (up to $J=130$) due to the heavy character
of the molecule, the size of the Hamiltonian matrices to be solved has
to be as large as 400,000$\times$400,000, which represents our biggest
calculation so far. This is despite the fact that only the smallest matrix
($A_1\p$ and $A_1\pp$ symmetries of \Dh{3}(M)) had to be considered due
to the nuclear spin statistics of $^{32}$S$^{16}$O$_3$. The sheer size
of these matrices requires special measures not only on the software
side (\trove), which is discussed below, but also from the hardware.
For this example of the 400K$\times$400K matrices, the
diagonalizations were performed on the Cambridge SMP facilities within
the DiRAC~II project, using about 1000 cores, 6 Tb of RAM and a
specially adapted version of the eigensolver PLASMA \cite{plasma}
by the SGI team for \trove.

In order to tackle these and other challenges associated with large basis sets
and matrix sizes, the following critical modifications of \trove\ have been
performed.

\textbf{Checkpointing.} The  production of a complete line list for a
polyatomic molecule with four and more atoms, takes very long times, longer
than the wall-clock limits of the high performance computers (HPC)
we have access to would allow. Therefore it was
important to implement the so-called `checkpointing' feature
(i.e. storing data required for a restart)
for all calculation steps of the
\trove\  protocol, together with a control
mechanism allowing a restart at any computational step. Moreover, the
eigen-coefficients in different representations used at different
stages are also stored in the form of `checkpoints' thus treating them on
the same footing. In order to prevent accidental usage of the wrong
checkpoints, most of these files contain a built-in structure of
`signatures' with a header containing some key parameters representing
a \trove\ project (expansion orders of the kinetic and potential
energy functions, sizes and types of the basis sets etc) and
control-phrases at the beginning and end of the different sections (e.g.
\verb!End Quantum numbers and energies!).

\textbf{Symmetries.} Each molecule type in \trove\ is represented as a project specifying reference (equilibrium) geometry, definition of the geometrically defined coordinates (GDC),  and description of the associated transformation properties of these coordinates as well as of the rigid-rotor wavefunctions used for the rotational basis set. The same molecule type allows different choices of GDC depending on the specifics of the system as well as of the reference geometries, with the reference configuration to be either rigid or non-rigid. The transformation symmetry properties of the rigid-rotor wavefunctions $\ket{J,K,\tau}$ \cite{05YuCaJe.NH3} will vary depending on the choice of the $z$-axis. For example in case of an XY\2\ molecule, the $z$-axis can be chosen along the bisecting vector or perpendicular to it, which changes the symmetry properties of $\ket{J,K,\tau}$. For most of the symmetries, the irreducible (symmetry-adapted) combinations of the rigid-rotor wavefunctions are obtained as Wang functions \cite{82Paxxxx.method}
\begin{eqnarray}
\label{eq:J,K,gamma:1}
\ket{J,K,\Gamma} &=& \frac{1}{\sqrt{2}} \left[ \ket{J,K} \pm \ket{J,-K}  \right] \quad (K\ne 0), \\
\label{eq:J,K,gamma:2}
\ket{J,0,\Gamma} &=& \ket{J,0}.
\end{eqnarray}
Because of this property and the fact that the Hamiltonnian operator
is quadratic in terms of the angular momentum operators $\hat{J}_x$,
$\hat{J}_y$, and $\hat{J}_z$, the ro-vibrational Hamiltonian matrix
has a block-diagonal structure with vanishing matrix elements for
$|K-K'|>2$.

In the course of the ExoMol project the following new molecular types and corresponding symmetries were implemented:   XY\4\ (\Td, \Cv{3}) \cite{jt555,jt572}, non-linear and non-rigid  X\2Y\2\ (\Cs(M), \Dh{2}(M), \Ch{2}$^{+}$(M), \Ch{2}(M), \Cv{2}(M)) \cite{jt553,jt620}, linear X\2Y\2\ (\Dh{n}(M), \Cv{2}(M), \Cs), rigid X\2Y\4\ (\Dh{2}(M)).

\textbf{Euler symmetry.}
The XY\4\ is a special case since the simple
symmetrization rules given in
Eqs.~(\ref{eq:J,K,gamma:1}) and (\ref{eq:J,K,gamma:2}) do not work
\cite{11AlLeCa.CH4} due to the additional symmetry axis (1,1,1)
required to define its equivalent rotations~\cite{99BuJexx.CH4}. The
adapted basis set is given in this case by a linear combination of
$\ket{J,k}$ basis functions with $k$ spanning different values ($-J \le k \le J$).
To address this problem a new routine for construction of symmetry-adapted rigid-rotor
wavefunctions has been implemented. This symmetrization approach is based
on the properties of the Wigner functions upon the
equivalent rotations of an arbitrary molecular symmetry group (not only \Td) and only requires
the values of the equivalent Euler angles $\alpha,\beta,\gamma$ only (see, for example, \cite{11AlLeCa.CH4}).
As a consequence abandoning the Wang-type structure from Eq.~(\ref{eq:J,K,gamma:1}) and (\ref{eq:J,K,gamma:2}),
the ro-vibrational Hamiltonian matrices do not have the $|K-K'|\le 2$ block-diagonal structure.
Furthermore,  matrix elements of the
dipole moment components in the laboratory-frame are also less compact. Therefore
the corresponding modules in \trove\ responsible for the ro-vibrational
Hamiltonian and dipole matrix elements had to be modified. More details can be
found in our paper presenting our very large hot  methane line
list known as 10to10. \cite{jt564}


\subsection{PES refinement}

The \trove-refinement method \cite{11YuBaTe.NH3} is based on the two main features: (i) the eigenfunctions of the ro-vibrational Hamiltonians (usually $J\le 5$) corresponding to the \ai\ potential energy function $V^{\rm ai}(\underline{Q})$ are used as basis functions to solve the \schr\ equations for the modified potential function  $V^{\rm R}(\underline{Q})$ during the refinement procedure; (ii) the refined potential energy function is represented  as a
correction $\Delta V(\underline{Q})$ to the \ai\ PES as given by
\begin{equation}
\label{e:fit-PES}
V^{\rm R}(\underline{Q}) = V^{\rm ai}(\underline{Q}) + \Delta V(\underline{Q}).
\end{equation}
 The refined part of the PES $V^{\rm R}(\underline{Q})$ is in turn represented as an expansion in terms of the internal coordinates
\begin{equation}
\label{e:fit-Delta}
\Delta V(\underline{Q}) = \sum_{ijk\ldots} \Delta F_{ijk\ldots } Q_1^i\, Q_2^j\, Q_3^k\, \cdots .
\end{equation}
with the expansion coefficients $\Delta F_{ijk\ldots }$
being varied using the Hellman-Feynmann theorem. The term $\Delta
V(\underline{Q})$ plays the role of the external function at Step~\ref{i:vib}.
The \trove\ refinement project requires the following additional calculation
steps after  Step~\ref{i:diag-rovib} in the calculation protocol above
\begin{enumerate}
 \setcounter{enumi}{10}
  \item The vibrational matrix elements of  $\Delta V(\underline{Q})$ (for a given approximation) are converted from the $J$$=$$0$ representation (Step~\ref{i:conv}) to the representation of the \ai\ ro-vibrational eigenvectors;
  \item At each iteration, a set of refined $H^{\rm R}= H^{\rm ai}+ \Delta V$ Hamiltonian matrices are constructed and diagonalized;
  \item The eigenvalues are compared to the experimental energy levels;
  \item The diagonal matrix elements of $Q_1^i Q_2^j Q_3^k \ldots$ on these eigenfunctions are computed and used to evaluate the next approximation for $\Delta V(\underline{Q})$;
  \item The fitting iteration steps are repeated until all accuracy or convergence criteria are satisfied.

\end{enumerate}

In order to prevent non-physical distortions of $V^{\rm
  R}(\underline{Q})$, the refinement is usually constrained to the original
\ai\ potential function. This is achieved by a simultaneous fit
\cite{03YuCaJe.PH3} of the potential parameters $\Delta
F_{i,j,k,\ldots}$ to the experimental energies and \ai\
potential function evaluated on a grid of (usually 10,000--20,000)
geometries. The current \trove\ implementation of the refinement approach given by Eqs.~(\ref{e:fit-PES}) assumes that the same functional form (not necessarily polynomial) in Eq.~(\ref{e:fit-Delta})  is also used for $V(\underline{Q})$:
\begin{equation}
\label{e:PES:expansion}
V(\underline{Q}) = \sum_{ijk\ldots} F_{ijk\ldots } Q_1^i\, Q_2^j\, Q_3^k\, \cdots .
\end{equation}
This simplifies the evaluation of the derivatives of diagonal  Hamiltonian matrix elements $\bra{i}H\ket{i}$ with respect to $\Delta F_{ijk\ldots} $ needed for the least-squares fit via the following property:
$$
\frac{\partial \bra{i}H\ket{i}}{\partial \Delta F_{ijk\ldots}} = \bra{i}H(F_{ijk\ldots}=1,F_{i'j'k'\ldots}=0)\ket{i},
$$
where $H(F_{ijk\ldots}=1,F_{i'j'k'\ldots}=0)$ denotes the Hamiltonian operator with all potential parameters $F_{i'j'k'\ldots}$ set to zero except $F_{ijk\ldots}$ which is set to one. That is, the same subroutine can be used to evaluate both $V(\underline{Q})$ and $\Delta V(\underline{Q})$. It should be noted however that an independent form for $\Delta V(\underline{Q})$ can be also easily implemented if required.

\subsection{Curvilinear coordinates}

Originally {\sc TROVE} was  based on the expansion in terms of
linearized coordinates around an equilibrium geometry in the case of a rigid
molecule or a one-dimensional non-rigid reference configuration
\cite{83Jensen.method} in the case of molecules with one large amplitude
motion (e.g. ammonia or hydrogen peroxide). Linearized coordinates
are defined as a linear expansion of GDCs in terms of the Cartesian
coordinates displacements truncated after the linear term \cite{05YuThCa.method}.
The linearized coordinates have the
advantage of simplifying the Eckart conditions \cite{TROVE}.  Very
recently, {\sc  TROVE} has been extended for expansions in
terms of geometrically defined (or curvilinear) coordinates. In order
to be able to use the Eckart conditions in this case, an automatic
differentiation (AD) procedure has been implemented \cite{15YaYuxx.method}.
Use of the curvilinear coordinates
significantly improves the basis set convergence\cite{15YaYuxx.method}.
This method was originally tested on
NH\3, PH\3, CH\3Cl and H\2CO, and has been used in subsequent
applications \cite{15OwYuYa.SiH4,jt612,jt634}. AD is a robust
numerical method to compute derivatives of arbitrary functions by
computer programs.

\subsection{Dipole moments}

As for the potential energy functions, at least in principle, \trove\
accepts any analytical form for the electric dipole moments used for
intensity calculations. This is because \trove\ re-expands any
use-defined function in terms of \trove\ internal coordinates (either
linearized \cite{TROVE} or curvilinear \cite{15YaYuxx.method}) using
numerical finite differences. \trove\ requires that the corresponding
subroutine outputs the dipole moment components for any given
instantaneous molecular geometry in Cartesian coordinates. Analytical
forms for the following dipole moment functions have been implemented
in \trove: XY\3-type molecular bond (MB) \citep{05YuCaLi.NH3} and
symmetrized MB representations \citep{09YuBaYa.NH3}; an HSOH-type
dipole moment function (DMF) \citep{09YuYaTh.HSOH}; an H$_2$CS-type
DMF \citep{13YaPoTh.H2CS,jt597}; an XY\4-type symmetrized MB
representation \citep{jt555}; an HOOH-type dipole moment
function \citep{jt638}. Since these functions are based on some
user-defined choice of the coordinate system, an interface to
transform this system to the \trove\ coordinates (Cartesian) is always
required.

\subsection{{\sc TROVE} for a linear molecule}

{\sc TROVE} was originally written to treat non-linear molecules only.
This meant that {\sc TROVE} was not capable of treating accurately
enough for practical spectroscopic applications molecules such as
water, which has a relatively low barrier to the linearity.  This has
been addressed in the most recent version of {\sc TROVE} by extending
it to the so-called $3$$\times$$N$-$5$ approach,
\cite{jt22} where the rotation of the molecule around
the molecular axis (e.g. $z$) is excluded from the set of the Euler
angles and combined with the set of the vibrational modes. Technically
this is done by describing the deformation of the linear geometry
(displacement angle) and its rotation about $z$ via a double
degenerate coordinate ($q_x$,$q_y$) representing projections of the
bond angle onto the body-fixed $xz$ and $yz$ planes. The linearized
coordinates in this cases are best suited for this 2D internal mode.
All kinetic energy terms corresponding to the $z$ component are simply
set to zero and thus excluded from the calculations. The construction
of the $3$$\times$$N$-$5$ Hamiltonian requires minimal modifications
of the $3$$\times$$N$-$6$ code. However the ro-vibrational basis set
in the product form $\ket{J,k}\ket{v,l}$ has to be constrained as
follows \cite{70Watson.method}
$$
  k = l \equiv \sum_i l_i \, ,
$$
where $k$ is the rotational quantum number (projection of the
rotational angular momentum on $z$), $l_i$ are the vibrational angular
momenta and $v$ is a generic vibrational quantum number. Thus the
vibrational basis set has to be constructed with $l$ as a `good'
quantum number.  Full
details of our
$3$$\times$$N$-$5$ approach will be reported elsewhere.

\subsection{Rotational energy clustering}

{\sc TROVE} has been used to study the effect of the rotational energy
clustering \cite{72DoWaxx.cluster} for the XY\3\ type molecules SbH\3,
BiH\3, PH\3, AsH\3, and SO\3\ \cite{09YuOvTh.SbH3,14UnYuTe.SO3}. In
order analyze the associated localization rotations,
\cite{96KoPaxx.cluster} the following modules were implemented: (i)
construction of classical rotational energy surfaces
\cite{84HaPaxx.cluster}, (ii) construction of rotational probability
density \cite{05YuThPa.PH3} and (iii) determination of axes characterized by stable localized rotations.

\subsection{Temperature averaging and matrix exponent expansion}

For an ensemble of molecules in thermal equilibrium at absolute temperature
$T$ the thermal average of a molecular property $P$ is given by
\begin{equation}\label{e:taver}
\langle P \rangle_T = \frac{1}{Q} \sum_i g_i \exp\left(-\frac{E_i}{kT}\right) \langle P \rangle_i,
\end{equation}
where $g_i$ is the degeneracy of the $i$th state with the energy $E_i$ relative
to the ground state energy, $k$ is the Boltzmann constant and $Q$ is the internal partition
function defined as
\begin{equation}\label{e:partfunc}
Q = \sum_i g_i \exp\left(-\frac{E_i}{kT}\right),
\end{equation}
and $\langle P \rangle_i$ is an expectation value of $P$ in a rovibrational state $i$
\begin{equation}\label{e:expectval}
\langle P \rangle_i = \langle \Phi_i | P | \Phi_i \rangle .
\end{equation}
A \trove\ module for computing thermal averaging of a general molecular function based on Eqs.~(\ref{e:taver}--\ref{e:expectval}) was implemented and applied to indirect nuclear spin-spin coupling constants and equilibrium structure of of ammonia \cite{10YaYuPa.NH3} and isotropic hyperfine coupling constant of methyl radical \cite{15AdYaYuJe.CH3}.
The calculations require the ro-vibrational
eigenvalues $E_i$ and eigenvectors $\Psi_i$ obtained by a time-consuming matrix diagonalization.
An alternative to this (also implemented as a part of this module) is an averaging technique based on the construction of the density matrix $\rho_{i,i'}$ obtained by expanding the matrix exponent of a Hamiltonian matrix as a Taylor series:
\begin{equation}\label{e:exp:rho}
\rho_{i,i^\prime} =  \frac{1}{Q} \bra{\phi_i}  \exp\left(-H/kT \right)\ket{ \phi_i'} =
\sum_{k\ge 0}\frac{1}{k!} \bra{\phi_i}  \left(- H/kT \right)^k\ket{ \phi_i'}
\end{equation}
in the representation of the basis functions.
This approach is based on the realization that
Eq.~\eqref{e:taver} represents the trace of a matrix product:
\begin{equation}\label{e:taver:P}
\langle P \rangle_T = {\rm tr} (\rho_{i,i} \langle P \rangle_i),
\end{equation}
involving the (diagonal) density matrix
\begin{equation}\label{e:taver:H}
  \rho_{i,i}  \equiv
  \frac{1}{Q}  \exp\left(-\frac{E_i}{kT}\right)   =
  \frac{1}{Q} \bra{\Phi_i} \exp\left(-H/kT \right)\ket{\Phi_i}.
\end{equation}
Since the trace does not depend on the choice of the representation
Eq.~\eqref{e:taver:H} is conveniently evaluated in the basis set representation.

\subsection{GAIN}

The longest part of the line list production is usually the intensity
calculations. The hot line lists of polyatomic molecules typically
require billions of transition dipoles (linestrengths or Einstein A coefficients)
to be computed. A calculation of a linestrength (as well as of an
Einstein coefficient) requires a matrix element of the molecular
space-fixed dipole moment for all {\sc TROVE} ro-vibrational
eigenfunctions subject to the selection rules and thresholds (see {\sc
  TROVE} protocol above), i.e. a vector-matrix-vector product, each of
which is relatively small in terms of the memory costs and fully
independent from other transitions. This makes it perfectly suitable
for the GPU architecture. We have modified the intensity part of {\sc TROVE}
to make it compatible for and efficient with GPUs. The new {\sc TROVE}
module and the underlying approach is called {\sc GAIN} \cite{jtGAIN}.
With small modifications  {\sc GAIN} could be adopted for other
variational programs. The gain in the calculation speed is from a factor
of 10 to
1000 depending on the type of GPU used.

\section{Larger molecules}

As part of the ExoMol project we have worked with one further nuclear motion code
{\sc AngMol} which was originally developed by Gribov and Pavlyuchko \cite{88GrPa.method}.
With Pavlyuchko we developed a hybrid variational -- perturbation theoretical method for
treating both vibrational and vibrational-rotational motion \cite{jt588} and  computing
spectra of large, hot systems efficiently \cite{jt603}. This methodology has been used
successfully to obtain a  line list for hot nitric acid (HNO$_3$) \cite{jt614}.
However, {\sc AngMol} has been developed in a highly specific manner. Rather than continuing
its development, our plan is to implement the hybrid procedure successfully tested in
{\sc AngMol} within {\sc TROVE}.

\section{Conclusions and future developments}

The codes {\sc Duo}, {\sc DVR3D} and {\sc TROVE} are all publicly
accessible via the CCPForge program depository
(https://ccpforge.cse.rl.ac.uk/), where each of them are available as
a separate project.

A number of developments of these codes are in progress or being
planned. In particular, we are just starting to extend the polyatomic
codes to include transitions between different electronic states and
hence to consider the vibronic transitions already considered by the
diatomic code {\sc Duo}. The calculation of all states up to
dissociation for strongly-bound triatomics has been possible with {\sc
  DVR3D} for some time \cite{jt100,jt132,jt230}; this leads to the
possibility that wave functions generated in such calculations can be
used for low-energy (or cold) reactive problems which occur just above
dissociation. This possibility is currently being explored
\cite{jt643}.  Another development in progress is the extension of
{\sc TROVE} for molecular dynamics in the presence of external time
dependent electric fields. For example,  recently the  {\sc TROVE}
has been extended to allow time-dependent solutions of
Schr\"{o}dinger equations for polyatomic molecules exposed by electric
fields of arbitrary shapes and polarizations \cite{15YaYuxx.HSOH},
where the flexibility of the ExoMol format \cite{jt631} is explored
for the transition dipole and polarizability moments required to
simulate the laser-driven molecular dynamics.
Updated versions of the
codes containing these extensions and others will be placed in the
CCPForge  program depository in due course.

\subsection*{ACKNOWLEDGMENTS}

This work was supported by the ERC under Advanced Investigator Project 267219.
We thank the other members of the ExoMol team for their participation
in the many program developments discussed in this article.

\bibliographystyle{apsrev}

\clearpage

\begin{figure}
\centering
\includegraphics[height=100mm]{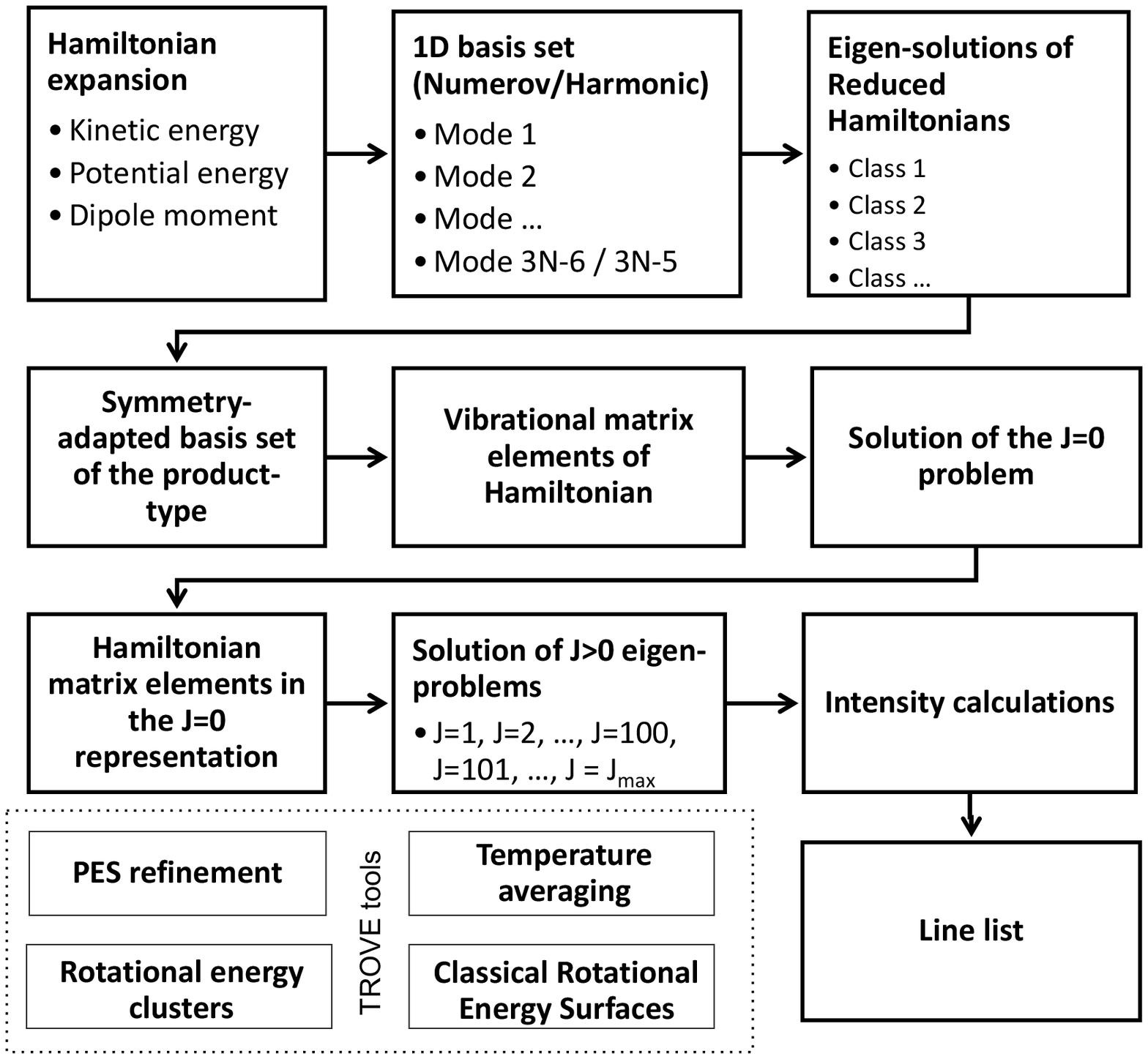}
\caption{\label{f:trove} Flow chart illustrating different stages of the \trove\ algorithm \cite{DVR3D,TROVE,15YaYuxx.method}.}
\end{figure}

\end{document}